\documentclass[twocolumn,doublespacing,showpacs,preprintnumbers,amsmath,amssymb]{revtex4}
\usepackage{amssymb}
\usepackage{txfonts}
\usepackage{graphicx}
\usepackage{subfigure}
\usepackage{dcolumn}
\usepackage{bm}
\usepackage{sidecap}

\begin{document}

\title{Improved Four-state Continuous-variable Quantum Key Distribution with Long Secure Distance}
\author{Jian Yang}
\author{Bingjie Xu}
\author{Xiang Peng}
\thanks{Corresponding author: xiangpeng@pku.edu.cn}
\author{Hong Guo}
\thanks{Corresponding author: hongguo@pku.edu.cn.}
\affiliation{CREAM Group, State Key Laboratory of Advanced Optical
Communication Systems and Networks (Peking University), 
School of Electronics Engineering and
Computer Science, Peking University, Beijing 100871, PR China }
\date{\today}
\begin{abstract}
The four-state continuous-variable quantum key distribution (CVQKD) protocol
has a long practical secure distance \cite{Leverrier_PRL_2009}, while
it has the difficulty of parameter estimation.
We propose an improved four-state protocol, where the covariance matrix can
be estimated from experimental data without using the linear channel assumption,
and thus ensuring its unconditional security in the asymptotical limit. Our new
scheme keeps the advantages of high reconciliation efficiency and long secure
distance of the four-state protocol, and it can be implemented under current
technology.
\end{abstract}

\maketitle



\section{Introduction}

Quantum key distribution (QKD) is one of the most practical
application of quantum information, which allows two remote
parties, Alice and Bob, to establish a sequence of   secure
keys \cite{Scarani_RMP_09}.  Continuous-variable
quantum key distribution (CVQKD) encodes        information
into the quadratures $x$ and $p$ of the optical field, 
and extracts it with homodyne
detections, which usually have higher repetition rate  than
that of single-photon detections. So, CVQKD can potentially
generate secure keys with higher speed. Historically, CVQKD
protocols are at first based on squeezed states
\cite{Hillery_PRA_2000, Cerf_PRA_2001}. Later,
coherent state protocols with Gaussian modulation      were
found to be more practical choices
\cite{Grosshans_PRL_02, Weedbrook_PRL_04}. Both   protocols
have been experimentally                       demonstrated
\cite{Grosshans_Nature_03, Lodewick_Ex_PRA_07}     and have
been shown secure against arbitrary collective      attacks
\cite{Grosshans_PRL_05, Navascues_PRL_05}, which        are
optimal in the asymptotical limit \cite{Renner_PRL_102}.

One
remaining problem is that the reconciliation     efficiency
$\beta$ is quite low for Gaussian modulation,    especially
when the transmission distance is long. As mentioned in
\cite{Leverrier_PRL_2009}, this is the main limiting factor
of the secure distance. There are two possible ways to solve this
problem. One is to build a good reconciliation code with
reasonable efficiency even at low SNR (signal to noise
ration), which has been achieved very recently
\cite{Jouguet_PRA_2011}. The other is to
use discrete modulation, such as the four-state protocol,
proposed by Leverrier \textit{et al.}
\cite{Leverrier_PRL_2009}. In this protocol,
Alice randomly prepares one of the four coherent states:
$|\alpha_{m}\rangle_{B}=|\alpha e^{i(2m+1)\pi/4}\rangle_{B}$
with $m\in \{0, 1, 2, 3\}$ and sends to Bob. Then, Bob
randomly measures the $x$ or $p$ quadrature of the signal
pulse as his result, the sign of which encodes the bit of
the raw key. Since the sign of quadrature has discrete
possible values, there exist very good error correction
codes when
extracting $I(a:b)$, even for extremely low SNR. From this
viewpoint, the four-state protocol combines the high
reconciliation efficiency of discrete modulation and the
security proof of CVQKD together, and improves the
secure distance effectively.

However, in this scheme, Alice and Bob can not estimate
the covariance matrix from their experimental data without
the linear channel assumption (LCA) in practice.
In the entanglement-based (E-B) scheme of
the four-state protocol, the projection measurement
$\{|\psi_{m}\rangle\langle\psi_{m}|, m=0, 1, 2, 3\}$ Alice
performs only helps
to discriminate which coherent state is sent to Bob, but
does not measure the quadratures of her mode. So, Alice and
Bob are not able to evaluate the covariance matrix
$\gamma_{AB}$ from experimental data unless using the LCA,
which compromises the security of the
protocol. To solve this problem, Leverrier \textit{et al.}
modified their protocols by introducing decoy states
\cite{Leverriver_PRA_2011}, such that
\begin{equation}
p\rho_{\rm key}+(1-p)\rho_{\rm decoy}=\rho_{\rm G},
\end{equation}
where $\rho_{\rm key}$ is the state sent to Bob in the
four-state protocol and $\rho_{\rm decoy}$ is the decoy state.
Alice randomly prepares $\rho_{\rm key}$ and $\rho_{\rm decoy}$
with probability $p$ and $1-p$, respectively, so that the
mixed state sent to Bob is  Gaussian, $\rho_{G}$. The main
difficulty of this method is the decoy state $\rho_{\rm decoy}$
can not be accurately prepared.

In this paper, we proposed an improved four-state protocol
by modifying its entanglement-based (E-B) scheme, the covariance
matrix of which can be directly evaluated from
experimental
data without using the LCA, and its
corresponding prepare and measurement (P\&M) scheme is not
difficult to implemented under current technology. Using
discrete coding, the high reconciliation efficiency and
long secure distance can be kept in this protocol.

\section{The Improved Entanglement-based Scheme of The Original Four-state Protocol}
In this section, we introduce the improved E-B
scheme of the original four-state protocol. In practice, CVQKD
protocols are implemented in the P\&M scheme, 
and the secure key rate against collective
attacks can be calculated by
\begin{equation}
K_{R}=\beta I(a:b)-S(b:E),
\end{equation}
where $K_{R}$ is the secure key rate using reverse
reconciliation,
$I(a:b)$ is the classical mutual information between
Alice and Bob,
$S(b:E)$ is the quantum mutual information between Bob
and Eve,
and $\beta$ is the reconciliation efficiency. $I(a:b)$
can be
directly estimated from experimental data,
while $S(b:E)$ should be estimated using its equivalent
E-B scheme. In the E-B scheme
of original four-state protocol \cite{Leverrier_PRL_2009},
Alice prepares
\begin{equation}\label{eq:Leverrier_EB}
|\Phi_{L}\rangle_{AB}=\frac{1}{2}\sum_{m=0}^{3}|\psi_{m}\rangle_{A}|\alpha_{m}\rangle_{B},
\end{equation}
measures mode $A$ with
$\{|\psi_{m}\rangle_{A}\langle\psi_{m}|\}$, and sends mode
$B$ to Bob,
where $\{|\psi_{m}\rangle_{A}\}$ are orthogonal states
and $m\in\{0, 1, 2, 3\}$.
As mentioned above, the main difficulty of this E-B
scheme is parameter estimation,
where Alice's measurement
$\{|\psi_{m}\rangle\langle\psi_{m}|\}$ only helps her
to discriminate which state is sent to Bob, but does not
provide any information about the quadratures of her mode.
Comparatively, in the E-B scheme of Gaussian modulation
protocols \cite{Grosshans_QIC_2003}, Alice
prepares EPR pairs, and measures her mode with heterodyne 
detection, which not only projects Bob's mode
into coherent states, but also provides the information
about the quadratures of Alice's mode, with which Alice
and Bob are able to estimate the covariance matrix
$\gamma_{AB}$ from their experimental data.

Our improvement is to
substitute $\{|\psi_{m}\rangle_{A}\}$ with proper states $\{|\psi'_{m}\rangle_{A}\}$.
Obviously, there are at least two conditions that
$|\Phi'\rangle_{AB}=\sum_{m=0}^{3}C_{m}|\psi'_{m}\rangle_{A}|\alpha_{m}\rangle_{B}$
should satisfy, where $C_{m}$ is the normalization coefficient:
\begin{enumerate}
  \item Alice's mode
$\{|\psi'_{m}\rangle_{A}\}$ can be discriminated by
homodyne or heterodyne detections, with which Alice are able to measure
the quadratures of mode $A$ and the covariance matrix $\gamma_{AB}$
can be estimated from experimental data.
  \item The covariance matrix of $|\Phi'\rangle$ should be as close to that of
Gaussian state as possible, which ensures the secure bound is tight, since the
Gaussian optimality theorem is used when calculating $S(b:E)$.
\end{enumerate}
From this viewpoint,
the original E-B model in Eq. (3) satisfies condition 2, since its covariance
matrix is close to that of EPR state, especially when the modulation is small.
Its main drawback is that Alice does not use homodyne or heterodyne detections, which
does not satisfies condition 1.

\begin{figure}[t]
\includegraphics[height=0.8 in]{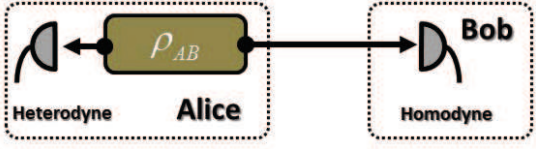}
\caption{(color online) Preliminary model for improving the E-B scheme. Alice
prepares mixed states $\rho_{AB}$, measures $x$ and $p$ of her mode, and sends
the other to Bob. Then, Bob measures the quadratures of his mode with homodyne
detection. Though this model can not generate secure keys directly, it is very
enlightening.
}\label{EB1}
\end{figure}

A natural choice for $|\psi'_{m}\rangle$ is coherent states,
$|\psi'_{m}\rangle=|\beta_{m}\rangle=|\beta e^{i(2m+1)\pi/4}\rangle$, where $\beta$ 
is real 
and $m=0, 1, 2, 3$.
When $\beta$ is large, states $\{|\beta_{m}\rangle\}$
can be discriminated by heterodyne detection
approximately. However, in this case,
Alice's measurement projects Bob's state $\rho'_{B}$ into a superposition
of coherent states, which is different from the $\rho_{B}$ of the 
original
four-state protocol, and its equivalent P\&M scheme is difficult to
implement in real experiment.

\subsection{The Mixed-state Scheme}
To avoid the problems above, we consider that Alice prepares
mixed state $\rho_{AB}$ and  measures $x$ and $p$ of
mode $A$ simultaneously with heterodyne detection, where
\begin{equation}
\rho_{AB}=\frac{1}{4}\sum_{m=0}^{3}|\beta_{m}\rangle_{A}\langle\beta_{m}|\otimes |\alpha_{m}\rangle_{B}\langle\alpha_{m}|.
\end{equation}
As illustrated in Fig. 1, Alice  projects mode $B$ into
a classical mixture of coherent states, which can be
implemented in its P\&M counterpart. Then, Bob
randomly measures $x$ or $p$ of mode $B$ with homodyne detection to
extract the information.
In this scheme, $\rho_{B}$ is identical to that of original four-state
protocol in Eq. (3) \cite{Leverrier_PRL_2009}.
To calculate $S(b:E)$, we recall that the covariance matrix $\gamma_{AB}$ of 
$\rho_{AB}$
is defined by
\begin{equation}
(\gamma_{AB})_{ij}={\rm Tr}[\rho_{AB}\{(\hat{r}_{i}-d_{i}),(\hat{r}_{j}-d_{j})\}],
\end{equation}
where the elements of displacement vector $d_{i}$ and $d_{j}$ are
$0$ in this scheme. Without difficult calculation, we find the covariance
matrix of
$\rho_{AB}$ has the form that
\begin{equation}
\left(
  \begin{array}{cc}
    V_{A}\mathbb{I} & C_{AB}\mathbb{I} \\
    C_{AB}\mathbb{I} & V_{B}\mathbb{I} \\
  \end{array}
\right),
\end{equation}
where $V_{A}$ and $V_{B}$ are the variances of mode $A$ and $B$, and
$C_{AB}$ are their correlations.
After channel transmission, the covariance matrix is changed to
\begin{equation}
\left(
  \begin{array}{cc}
    V_{A}\mathbb{I} & \sqrt{\eta} C_{AB}\mathbb{I} \\
    \sqrt{\eta}C_{AB}\mathbb{I} & \eta(V_{B}+\chi)\mathbb{I} \\
  \end{array}
\right),
\end{equation}
where $\eta$ and $\chi=(1-\eta)/\eta+\epsilon$ are the channel
parameters. Both parameters can be estimated from experimental
data. Here, for the ease of theoretical research, we suppose
the channel is linear, and $\eta$ and $\epsilon$ are the
transmittance and excess noise, respectively. It should
be emphasized that this assumption is just for simplifying
the simulation, but not necessary in this scheme \cite{YujieShen__PRA_11}.

The classical mutual information $I(a:b)$ can be
calculated by  $1-H(e)$, where $e$ is the bit error rate and
$H(e)$ is the Shannon entropy. The calculation of $S(b:E)$ is
a little more complex. To maximize Eve's information,
Eve is supposed to purify the whole system $\rho_{AB}$ and the
quantum mutual information $S(b:E)$ is calculated by
\begin{equation}
S(b:E)=S(E)-S(E|b)=S(AB)-S(A|b),
\end{equation}
where $S(AB)$ and $S(A|b)$ can be derived from $\gamma_{AB}$,
using the Gaussian optimality theorem \cite{Patron_PRL_06, Navascues_PRL_06}.

It is not surprising that we can not acquire positive secure
key rate $K_{R}$ with this E-B scheme, and there are two reasons.
First, when calculating $S(b:E)$, we suppose Eve is able to
purify the whole system to maximize the information leaked to her.
So, this scheme just overestimates Eve's information, 
since $\rho_{AB}$ is initially in a mixed state, and
it is not difficult to find that $S(b:E)>0$, even
if the transmittance $\eta$ is $1$. Second,  the secure key rate 
in E-B scheme is related to how much
pure entangled pairs can be extracted from $\rho_{AB}$, while in this
E-B scheme, $\rho_{AB}$ is separable and contains
little entanglement. Though Alice and Bob are classically 
correlated, they can not distill secret information from experimental data. 
Nevertheless, this attempt is very enlightening for our
improved E-B scheme in the following.

\subsection{The Improved Entanglement-based Scheme}

In this subsection, we proposed our improved E-B scheme, which is
illustrated in Fig. 2.
Instead of mixed state $\rho_{AB}$, Alice prepares
four-mode pure state $|\Psi_{I}\rangle_{FGAB}$,
where $I$ denotes the improved E-B scheme and its subsystem $AB$
is identical to the mixed state $\rho_{AB}$ in Eq. (4), where
$${\rm Tr}_{FG}\{|\Psi_{I}\rangle_{FGAB}\langle\Psi_{I}|\}=\rho_{AB}.$$
The reason why we introduce two ancilla modes $FG$ is to
guarantee that the pure state $|\Psi_{I}\rangle_{FGAB}$ does exist.
In this scheme, modes $F$ and $G$ are used as neutral parties, the information
of which is controlled neither by Eve nor by Alice and Bob.
Alice measures $x$ and $p$ of mode $A$ simultaneously with
heterodyne detection, and then sends mode $B$ to Bob. It is not difficult
to verify that $\rho_{B}$ in this case is identical to that of
the original four-state protocol.
The classical mutual information $I(a:b)$ can be directly
calculated by $1-H(e)$, while Eve's
knowledge about Bob's data $S(b:E)$ depends on the covariance matrix
of $|\Psi_{I}\rangle_{FGAB}$. Certainly, we can derive the exact expression
of $|\Psi_{I}\rangle_{FGAB}$ and calculate its covariance
matrix, while using our previous technique \cite{YujieShen__PRA_11},
we find that this work is not necessary.
\begin{figure}[t]
\includegraphics[height=0.9 in]{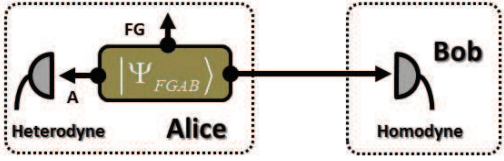}
\caption{(color online) The improved entanglement-based scheme. Alice
prepares pure states $|\Psi\rangle_{FGAB}$, measures $x$ and $p$ of
mode $A$ with homodyne
detections and sends mode $B$ to Bob.}\label{EB2}
\end{figure}

To compute $S(b:E)$, we consider a pure state
\begin{equation}
|\Psi_{L}\rangle_{FGAB}=|0\rangle_{F}|0\rangle_{G}|\Phi_{L}\rangle_{AB},
\end{equation}
where $|0\rangle$ is the vacuum state and $|\Phi_{L}\rangle_{AB}$ is
Leverrier's E-B model in Eq. (3).
It is not difficult to verify the covariance matrix of
$|\Psi_{L}\rangle_{FGAB}$ can be written as
\begin{equation}
\gamma'_{FGAB}=
\left(
  \begin{array}{cccc}
    \mathbb{I} & 0 & 0 & 0\\
    0 & \mathbb{I} & 0 & 0\\
    0 & 0 & V_{B}\mathbb{I} & Z\sigma_{z}\\
    0 & 0 & Z\sigma_{z} & V_{B}\mathbb{I}
  \end{array},
\right)
\end{equation}
where $Z$ is the correlation between Alice and Bob's quadratures \cite{Leverrier_PRL_2009}.
As shown in \cite{Nielsen2000}, since
$|\Psi_{I}\rangle_{AB}$ and $|\Psi_{L}\rangle_{AB}$ are different purifications of
$\rho_{B}$, there exist a unitary transformation $U_{FGA}$ on
mode $F$, $G$, and $A$, that
\begin{equation}
|\Psi_{I}\rangle_{FGAB}=U_{FGA}|\Psi_{L}\rangle_{FGAB},
\end{equation}
which does not change the mutual information $S(b:E)$,
since $U_{FGA}$ is commuted with $U_{BE}$, where $U_{BE}$
denotes Eve's operation on mode $B$ and $E$.
So, we can safely calculate $S(b:E)$ by substituting
$|\Psi_{I}\rangle_{FGAB}$ with $|\Psi_{L}\rangle_{FGAB}$,
the elements of which are known. This result can also be
understood physically. In reverse reconciliation, Both
Alice and Eve performs error correction according to Bob's
data. Whenever $|\Psi_{I}\rangle$ or $|\Psi_{L}\rangle$
is used, the mode $B$ sent to Bob is in the same state
$\rho_{B}$. Since Eve is not able to discriminate which 
E-B source is used, she has to perform the same 
strategy to eavesdrop the information, and  
the leaked information $S(b:E)$ should be same.

Since $|\Psi_{L}\rangle_{FGAB}$ is a pure state, we have
\begin{eqnarray}
S(E:b)&=&S(E)-S(E|b)\\
&=&S(FGAB)-S(FGA|b)\nonumber \\
&=&S(AB)-S(A|b),\nonumber
\end{eqnarray}
where van Neumann entropies $S(AB)$ and $S(A|b)$ can be calculated with
the symplectic eigenvalues of covariance matrices $\gamma_{AB}$ and $\gamma^{b}_{A}$
\cite{Garcia_PHD_2009}.

The performance of our improved E-B scheme is illustrated in Fig. 3, where
we use $\alpha=0.5$ and $\beta=20$. For small $\alpha$, the CM of
$|\Phi_{L}\rangle$ is close to that of EPR state, which ensures a high secure key
rate. For large $\beta$, coherent states $|\beta_{m}\rangle$ are approximately
orthogonal to each other, which are easier to be discriminated by heterodyne
detection. The variance of excess noise is set to be $0.002$, $0.004$, $0.006$,
$0.008$ and $0.01$, respectively. The secure distance is a little shorter than
that of original four-state scheme, and this is mainly because coherent states
$\{|\beta_{m}\rangle\}$ can not be discriminated deterministically.

\begin{figure}[t]
\includegraphics[height=2.3 in]{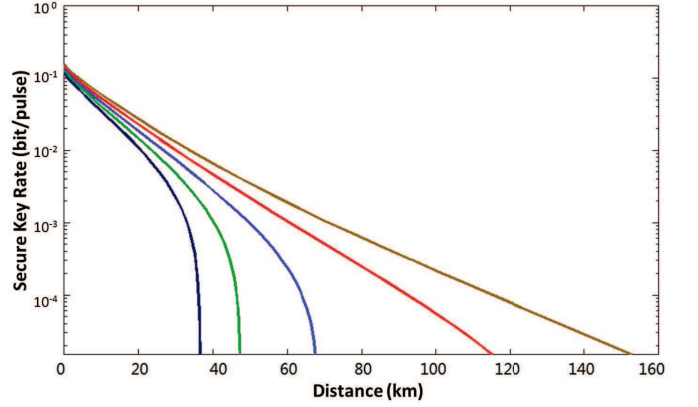}
\caption{(color online) The secure key rate of improved E-B scheme. The lines
from bottom to top correspond to $\epsilon=0.002, 0.004, 0.006, 0.008$ and $0.01$,
respectively.}\label{SK2}
\end{figure}

The reason why $\rho_{AB}$'s purification $|\Psi_{I}\rangle_{FGAB}$ can be 
used to generate secure keys is based on two sides.

First, $|\Psi_{I}\rangle_{FGAB}$ is a pure state, and Eve is not
benefit from her purification at the
very beginning. It is not difficult to
verify that $S(b:E)=0$, when $\eta=1$. Second, though $\rho_{AB}$ 
contains little entanglement, the whole system $FGAB$ is generally
an entangled-state, which can be used to extract secure keys.
From this viewpoint, our
improved E-B scheme just combines $\rho_{AB}$'s advantages in parameter
estimation and  $|\Phi_{L}\rangle$'s advantages in computing $S(b:E)$
together, which ensures a long secure distance.

\section{The Prepare and Measurement Scheme}
Though the E-B scheme is convenient for theoretical research,
it is difficult to implement directly. In this section,
we will present its equivalent P\&M scheme. As mentioned
above, in the E-B
scheme, Alice measures quadratures $x$ and $p$ of mode $A$
simultaneously. To do this, Alice should use a $50:50$
beamsplitter to separate mode $A$ into two parts, $A_{1}$ and
$A_{2}$, and the whole state is changed to
\begin{equation}
\rho_{A_{1}A_{2}B}=\frac{1}{4}\sum_{m=0}^{3}
|\frac{\beta_{m}}{\sqrt{2}}\rangle_{A1}\langle\frac{\beta_{m}}{\sqrt{2}}|\otimes
|\frac{\beta_{m}}{\sqrt{2}}\rangle_{A2}\langle\frac{\beta_{m}}{\sqrt{2}}|\otimes
|\alpha_{m}\rangle_{B}\langle\alpha_{m}|.
\end{equation}
Then, Alice measures $x$ of mode $A_{1}$, measures $p$ of mode $A_{2}$,
and projects Bob's state to
\begin{equation}
\rho_{B}|_{x_{A},p_{A}}=\frac{1}{4}\sum_{m=0}^{3}C_{m}^{(x_{A},p_{A})}|\alpha_{m}\rangle\langle\alpha_{m}|.
\end{equation}
The coefficient $C_{m}^{(x_{A},p_{A})}$ is calculated by
$$C_{m}^{(x_{A},p_{A})}=
{\rm tr(M_{A_{2}}(p_{A})M_{A_{1}}(x_{A})\rho_{A_{1}A_{2}}(m)M^{\dag}_{A_{1}}(x_{A})M^{\dag}_{A_{2}}(p_{A}))},$$
where operators
$M_{A_{1}}(x_{A})=|x_{A}\rangle_{A_{1}}\langle x_{A}|$,
$M_{A_{2}}(p_{A})=|p_{A}\rangle_{A_{2}}\langle p_{A}|$,
and $\rho_{A_{1}A_{2}}(m)=|\frac{\beta_{m}}{\sqrt{2}}\rangle_{A1}\langle\frac{\beta_{m}}{\sqrt{2}}|\otimes
|\frac{\beta_{m}}{\sqrt{2}}\rangle_{A2}\langle\frac{\beta_{m}}{\sqrt{2}}|$.
This
is a classical mixture of coherent states $\{|\alpha_{m}\rangle\langle\alpha_{m}|\}$,
where the probability $C_{m}^{(x_{A},p_{A})}$ is a Gaussian function
of Alice's measurement result $(x_{A},p_{A})$.
The calculation of $C_{m}^{(x_{A},p_{A})}$ is straight with the
methods in \cite{Leonhardt_1997}, while we omit the detail here and
and focus on its experimental realization. In this section,
we propose two possible P\&M schemes to implement this protocol. One is
the true random number generator (TRNG) based scheme and the other is
the beamsplitter based scheme.

\begin{figure}[t]
\includegraphics[height=0.9 in]{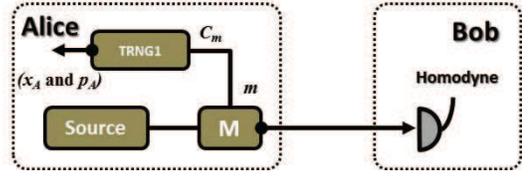}
\caption{(color online) The TRNG based scheme of improved
four state protocol. Alice generates random numbers $x_{A}$
and $p_{A}$ with TRNG1, and calculates
$\{C_{m}^{(x_{A},p_{A})}\}$, based on which Alice randomly
prepares $|\alpha_{m}\rangle_{B}$ to generates
$\rho_{B}|_{x_{A},p_{A}}$.
Bob extracts the information by randomly measuring $x$ or
$p$ of mode $B$ with homodyne detection.
} \label{PM1}
\end{figure}

\subsection{TRNG based Scheme}
In TRNG-based scheme, each time Alice uses TRNG1 to generate random pairs $(x_{A},p_{A})$
with probability
$${\rm Pr}(x_{A})={\rm tr}(M_{A_{1}}(x_{A})\rho_{A_{1}}M^{\dag}_{A_{1}}(x_{A}))$$
and
$${\rm Pr}(p_{A})={\rm tr}(M_{A_{2}}(p_{A})\rho_{A_{2}}M^{\dag}_{A_{2}}(p_{A})),$$
respectively, where density operators
$\rho_{A_{1}}={\rm tr}_{A_{2}B}(\rho_{A_{1}A_{2}B})$ and
$\rho_{A_{2}}={\rm tr}_{A_{1}B}(\rho_{A_{1}A_{2}B})$.
To prepare $\rho_{B}|_{x_{A},p_{A}}$, Alice randomly prepares a
coherent state $|\alpha_{m}\rangle_{B}$ from
$\{|\alpha_{m}\rangle_{B}, m=0, 1, 2,3 \}$ with
probability $C_{m}^{(x_{A},p_{A})}$ and sends it to Bob.
As illustrated in Fig. 4,
the TRNG-based scheme can be realized within current technology,
while it is still a little complicated, since each time two 
random numbers are generated
and the probability $\{C_{m}^{(x_{A},p_{A})}\}$ depends on the 
random pair $(x_{A},p_{A})$.

\subsection{Beamsplitter based Scheme}
To simplify the experimental implementation, we propose a 
beamsplitter-based scheme. Noticing that
$\rho_{AB}=\frac{1}{4}\sum_{m=0}^{3}|\beta_{m}\rangle_{A}\langle\beta_{m}|\otimes |\alpha_{m}\rangle_{B}\langle\alpha_{m}|$,
we find it can be directly implemented with a
beamsplitter. As illustrated in Fig. 5, Alice prepares a coherent state $|\gamma\rangle$,
and modulates it with a phase modulator, driven by TRNG2. Then, the modulated coherent
state is separated by a beamsplitter, the output states of which are $|\beta_{m}\rangle_{A}$ and
$|\alpha_{m}\rangle_{B}$, respectively.
Then Alice measures the $x$ and $p$ of mode $A$ simultaneously, and sends mode $B$ to Bob.
In this scheme, TRNG2 generates only 4 possible values $\{m=0, 1, 2, 3\}$,
which are easier to implement than  TRNG1 in Fig. 4.

\begin{figure}[t]
\includegraphics[height=1in]{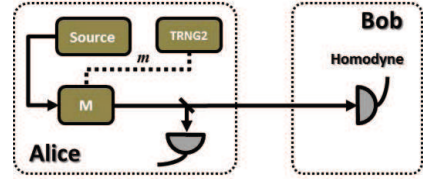}
\caption{(color online) The Beamsplitter based scheme of improved
four state protocol. Alice modulates coherent state with a phase
modulator driven by TRNG2 with $m=0, 1, 2, 3, {\rm and} \ 4$,
and separates it into two parts with a beamsplitter. Alice measures
the quadratures $x$ and $p$ of one output state and sends the other
to Bob.
} \label{PM2}
\end{figure}

\section{Discussion and Conclusion}

To sum up,
we propose an improved long-distance CVQKD protocol by
modifying the E-B scheme.
We find that the mixed state $\rho_{AB}$ is helpful
to establish a classical correlation between Alice and Bob,
and then purifies $\rho_{AB}$ with two ancilla mode $F$ and $G$
as the improved E-B model, $|\Psi_{I}\rangle_{FGAB}$.
The parameter estimation
can be performed in this
scheme without using the LNA.
Further, based on \cite{YujieShen__PRA_11},
we find the mutual information $S(b:E)$ of our
improved scheme is identical to that of Leverrier's one,
so we can derive a  security
bound for reverse reconciliation, without deriving the
exact expression of the ancilla states $F$ and $G$. From
this viewpoint, our improved E-B scheme combines the high
reconciliation efficiency of discrete coding and the facility of
parameter estimation together, and hence ensures a
long secure distance with unconditional security.
Also, we present two potential equivalent P\&M schemes
to implement the improved protocol experimentally.

There are also several remaining problems to study.
First, the
E-B model $|\Psi_{I}\rangle_{FGAB}$  can be further
optimized to make its density matrix closer to that of a EPR,
which may further improve the secure key rate. Second,
in four-state protocol, the optimal value of $\alpha$ is less
than 1, which is still not easy to detect by homodyne
detections in the experiment. At last, its unconditional
security against coherent attack need to be reconsidered when
the finite size effect is taken into account.

We thank Junhui Li for fruitful discussion.
This work is supported by the Key Project of National Natural
Science Foundation of China (Grant No. 60837004 and No. 61101081),
National Hi-Tech Research and Development (863) Program.

\end{document}